\documentclass{pnastwo}

\usepackage[pdftex]{graphicx}
\usepackage{xr}

\usepackage{amssymb,amsfonts,amsmath}

\begin{document}

\footcomment{Abbreviations: BP, Belief Propagation; MSTT, Minimum Steiner Tree; PCST, 
Prize Collecting Minimum Steiner Tree}

\title{Finding undetected protein associations in cell signaling by belief propagation}

\author{
M. Bailly-Bechet\affil{1}{LBBE, CNRS UMR 5558, Universit\'e Lyon 1, Villeurbanne, France}, 
C. Borgs\affil{2}{Microsoft Research New England, One Memorial Drive, Cambridge, USA},
A. Braunstein\affil{3}{Human Genetics Foundation, Via Nizza 52, 10126 Torino, Italy \& Politecnico di Torino, C.so Duca degli Abruzzi 24, Torino, Italy},
J. Chayes\affil{2}{}, 
A. Dagkessamanskaia\affil{4}{Universit\'e of Toulouse, UMR CNRS-INSA 5504  \& INRA 792, Toulouse, France}, 
 J.-M. Fran\c{c}ois\affil{4}{}\and  R. Zecchina\affil{5}{Politecnico di Torino, C.so Duca degli Abruzzi 24, Torino, Italy \& Human Genetics Foundation, Via Nizza 52, 10126 Torino, Italy}}

\contributor{Submitted to Proceedings of the National Academy of Sciences 
of the United States of America}

\maketitle

\begin{article}

\begin{abstract}
External information propagates in the cell mainly through signaling cascades
and transcriptional activation, allowing it to react to a wide spectrum of
environmental changes. High-throughput experiments identify numerous molecular
components of such cascades which however may interact through other undetected
molecules.

Some of these interactions may be detected using data coming from the
integration of a protein-protein interaction network and mRNA expression
profiles. This inference problem can be mapped onto the one of
finding appropriate optimal connected sub-graphs of a network defined by these
datasets. The optimization procedure turns out to be computationally
intractable in general. Here we present a new distributed algorithm for this
task, inspired from statistical physics, and apply this scheme to alpha factor
and drug perturbations data in yeast. We identify the role of the COS8 protein,
a member of a gene family of previously unknown function, and validate the
results by genetic experiments.

The algorithm we present is specially suited for very large data
sets, can run in parallel and can be adapted to other problems in systems
biology. On renowned benchmarks it outperforms other algorithms in the field.

\end{abstract}

\keywords{COMPUTATIONAL BIOLOGY | BELIEF PROPAGATION |  STEINER TREES | PHEROMONE PATHWAY}

\section{Introduction}
\dropcap{S}ignaling cascades, an exemplar of which is the phosphorylation
MAPK kinase pathways, consist in sequential reactions starting at receptor
proteins and transmitted through protein interactions to effector proteins.
Activation of these effectors leads to cellular changes, notably at the
transcriptional level, and results in the adaptation of the cell to its
surroundings \cite{Elston:2008xv}.  Identifying signaling pathways  is
particularly important for medical studies since their 
malfunction is responsible for many diseases, such as cancer \cite{King2006},
or Alzheimer \cite{Pei2008}.

From an engineering point of view, signaling cascades present interesting
properties: they provide signal filtering  \cite{Thattai:2002hz} and
amplification  \cite{Kholodenko:2006iz}. Their global interconnected
organization equip the cell with an integrated sensor network where pathways
can modulate one another through cross-talk and retroactions. In this complex
system, signal specificity is maintained by scaffold proteins
\cite{Locasale2008,Bashor:2008lk} acting as connectors of particular reactions.
Finally, the output of the information carried by the transduction network
allows for another layer of regulation,  namely combinatorial control in gene
expression \cite{Benayoun2009a}.
A purely experimental approach to the identification of  all components of a
pathway, or all components of a functional gene module, would need long and
costly experiments.  Such  process   would greatly  benefit from the extraction
of indirect information about pathways  from producible large-scale data, such
as  expression, sequencing  and protein interaction data. 
Indeed, even if the correlation between signaling pathway activity and
expression data may be weak (although this may be case-dependent, see 
\cite{Soufi2009} as an interesting
example), expression data are available in shear quantity. By introducing a
parameter weighting the relative importance given to expression data with
respect to reliable protein-protein interaction data or in general established
pathways knowledge,  we hope to be able to enrich the existing knowledge by the
small amount of information needed to reveal unknown interactions.  To this
scope, important aspects like the varying reliability of interaction data and
the proliferation of alternative paths which need to be compared, require the
development of heuristic algorithmic techniques which need to be efficient on
large scale data sets.

Here we propose a new method for the inference of hidden
components of functional networks and signaling pathways, from large-scale
transcriptomics and protein interaction data. Such functional networks, composed of
proteins acting together in given environmental conditions,  are an integrated
way of describing information processes in the cell.  This problem has enormous
potential applications and has already been addressed in several works
\cite{Scott:2005ao,Scott2006,White2008,Zhao:2008lr}, leading to interesting
theoretical predictions. In these works, the underlying methodology consists in
separately identifying single signaling pathways, and then collecting them in
an aggregated network. The methodology proposed here attempts instead to extract
information on an entire network, defined as a connected sub-graph of the full
protein interaction network. 
This technique was roughly sketched in a biological inference
context\cite{MBB:CMSB}. Here we present a complete, in-depth description and
analysis of the approach, including in particular algorithmic and experimental
validations, and comparison with results from a previous work along the same
lines. We also give full details of the algorithmic framework we use, which
may allow implementation of the same ideas to similar systems biology problems.

We state the functional network inference
problem in a rather simple and general graphical form.
Given a graph $G=(V,E)$ -- the protein interaction network (PIN) --  with
positive costs over edges $\left\{ c_{e}:e\in E\right\}$
 and positive prizes over vertices $\left\{ b_{i}:i\in V\right\}$, find a
 connected sub-graph $G'=(V',E')$ that minimizes the following function:
\begin{equation}
\min_{\begin{array}{c}
E'\subseteq E,V'\subseteq V\\
(E',V')\,\mbox{connected}\end{array}}\sum_{e\in E'}c_{e}-\lambda\sum_{i\in V'}b_{i}
\label{eq:H}
\end{equation}

To our purpose the costs of edges $c_e$ are chosen so that high confidence
interactions (protein interactions verified in small-scale experiments or found
in many large-scale datasets) have lower value with respect to low confidence
ones (interactions experimentally shown only once in a large-scale experiment).
The node prizes are computed by  $b_{i}=- \log p_i$, where $p_i$ is the
$p$-value of differential expression of node $i$ in the corresponding
microarray. The parameter $\lambda$ regulates the tradeoff between the edge
costs and vertices prizes, and its value indirectly controls the size of the
sub-graph $G'$. 

In spite of its apparent simplicity, the problem of solving Eq.~[\ref{eq:H}],
known as \emph{Prize-collecting Steiner Tree} problem (PCST), is computationally
intractable (NP-Hard) and heuristic algorithms need to be developed in order to
solve instances arising from large datasets. Satisfying the
connectivity constraint on the optimization task constitutes a major
computational difficulty. The problem remains intractable even in the case in
which $b_{i}\in\{0,L\}$ for a large $L>0$, as this limit case
corresponds to the better known \emph{Minimum Steiner Tree} problem (MSTT) on
graphs, which is also NP-Hard.

Modeling ideas related to our work are discussed in
\cite{Klau-pcst-2006,Klau-isbm-2008,Fraenkel2009, Yosef2009}. These studies
rely on different algorithmic techniques, namely on a combination of linear
programming relaxation solvers, branch and bounds optimization methods and
preprocessing of the underlying biological network. A detailed comparison shows
essentially that the difference in performance between LP-based methods and the
one presented here increases dramatically with the problem size. Additionally,
the computational cost of our approach scales roughly linearly with the size of
the problem and the algorithm is fully parallelizable. These two facts suggest
that the method proposed here may be particularly well-suited to study problems
defined on large networks.

The paper is organized as follows: First we present the general problem of
identifying optimal subgraphs as a technique for integrating different data
types. Then we discuss a new algorithmic approach based on belief propagation:
we provide benchmark performances together with a specific  application to
pheromone response data in yeast. Finally we describe in detail the
experimental validation of the predictions relative to the functional role of a
family of genes (COS). Complete details are given in the Supporting Information.

\section{Results}
\label{sec:pm1}

\subsection{A Message-Passing algorithm for PCST}

\begin{figure}
\begin{center}
\includegraphics[width=1.0\columnwidth]{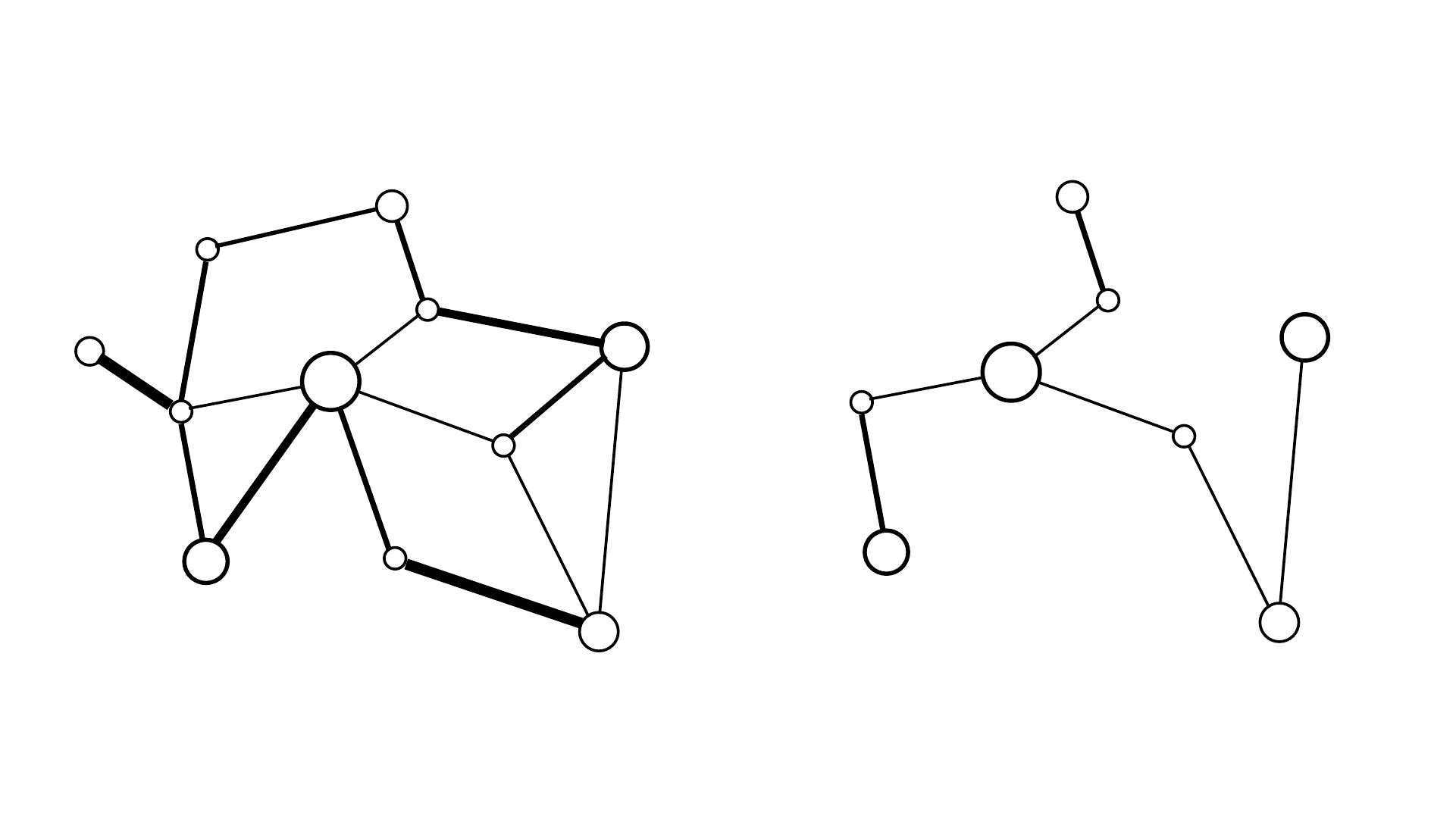}
\end{center}
\caption{An example of a prize-collecting Steiner tree. Larger nodes mean
larger prizes, thickness of the edges is proportional to their cost. A prize
collecting Minimum Steiner Tree (right) picks as many as possible of the larger
nodes while simultaneously picking the thinnest links and maintaining
connectivity. The analyzed yeast protein network has approximately 5000 nodes
and 22,000 edges\label{fig:model}}
\end{figure}

In \cite{BBZ2008a}, a statistical physics analysis of the 
properties of Steiner Trees on different ensembles of large random graphs. Here
we generalize this work and introduce a new algorithm which can be used to
identify signaling pathways in transduction PINs.  A detailed discussion is
given in the Supporting Information (SI).  Minimizing equation (\ref{eq:H})
gives access to {\it connected} networks which include reliable edges and, at
the same time, nodes which are significantly differentially expressed (see
Fig~\ref{fig:model}). This cost function could easily be generalized to other
type of interactions, e.g. 
gene-based information such as results of knock-out experiments. Biological
priors such as the relative position of proteins (nodes) along the tree or their
expected degree  could also be easily included in the same scheme.

One main difficulty with an optimization over connected sub-graphs is that the
connectivity condition is global rather than local, i.e. \emph{can not} be
verified by a set of simple local checks over the graph. This problem is dealt
with here by switching to a richer description of the sub-graph that includes
an extra variable for each graph node, which essentially denotes when (if ever)
nodes would be visited by an algorithm that explores the sub-graph from a given
starting ``root'' node. While such representations are in one-to-one
correspondence to connected sub-graphs, the connectivity condition \emph{does}
have an expression as a set of simple local conditions for the new variables.

The proposed algorithm consists in a set of functional equations for estimating
the probabilities that individual links belong to the optimal sub-graph
at a given distance from the starting node. Such equations can be written in a
computationally efficient form that is solved by iteration in a 
so--called message--passing procedure.  The general derivation and other
details are given in  SI text, whereas the source code is available for
download \cite{cmp}. The proof that in certain limit cases the algorithm
provides optimal results can be found in \cite{BBZ-JMP-08}.

\subsection{Tests and data analysis}
In order to assess the general efficacy of the algorithm, it was tested against
the collection of MSTT benchmark problems in the SteinLib data set \cite{KMV00}
which defines the state of the art in the field. Though the benchmarks problems
are not of biological nature, they are both large and difficult to solve, and
therefore particularly useful for  comparing performance of different
algorithms.  Quite surprisingly the  best known cost of  almost all  of the  open
problems could be improved in a fraction of the computational time (details 
and complete Tables in SI text).  Most of
the heuristic algorithms with   the previously best known performance  are
based on Linear Programming (LP) relaxations complemented with preprocessing of
the underlying graph and a branch-and-bound strategy, e.g.
\cite{Klau-pcst-2006} (details also in SI). Further direct comparisons between
such methods  and our approach suggest that already for moderate size networks
the LP-based methods become highly inefficient (details in SI).

As a second preliminary test, on biological data, we have compared our
technique to another method for the inference of linear signaling pathways
based on color-coding \cite{Scott2006}, the optimization criterion of which is
a restriction of ours to the edge cost part. We have assessed our algorithm
performance relative to \cite{Scott2006} with the same data and optimization
criterion. Essentially, the same pathways were found much more efficiently (due
to the fact that the computational cost does grow only linearly with the chain
length and not exponentially as in \cite{Scott2006}), and it was possible to
recover their variability by adjusting the chain length (details in SI text),
thereby proving the capacity of our algorithm to recover known biology.

\subsection{Pheromone response data}

Finally, we have applied the algorithm to analyze pheromone sensing on a yeast
protein network built by fusion of the MIPS\cite{Gueldener2006} and
DIP\cite{Xenarios:2000sy} networks, using 56 large-scale expression datasets
created to reconstruct the pheromone pathway experimentally by studying the
expression of strains deleted for key genes in the pathway \cite{Roberts2000}.
This system was chosen as a case study in virtue of a  pre-existing good
theoretical understanding of its functioning. The pheromone response system is
in fact a well-studied MAPK kinase cascade, which permits communication
previous to mating in haploid yeasts. Upon sensing a pheromone, cell cycle is
arrested, cytoskeleton and membrane structure are modified, and finally mating
occurs by fusion of two cells of opposite sexual type. 

The identification of optimal sub-networks was done for a range of $\lambda$
values, giving us variable structures going from the backbone of the network to
a very detailed picture of each sub-pathway. The root of the trees was defined
to be STE2, the $\alpha$-pheromone receptor and entrance of the entire
pheromone pathway \cite{Jenness1983, Burkholder1985}.

\begin{figure}
\includegraphics[width=1.0\textwidth]{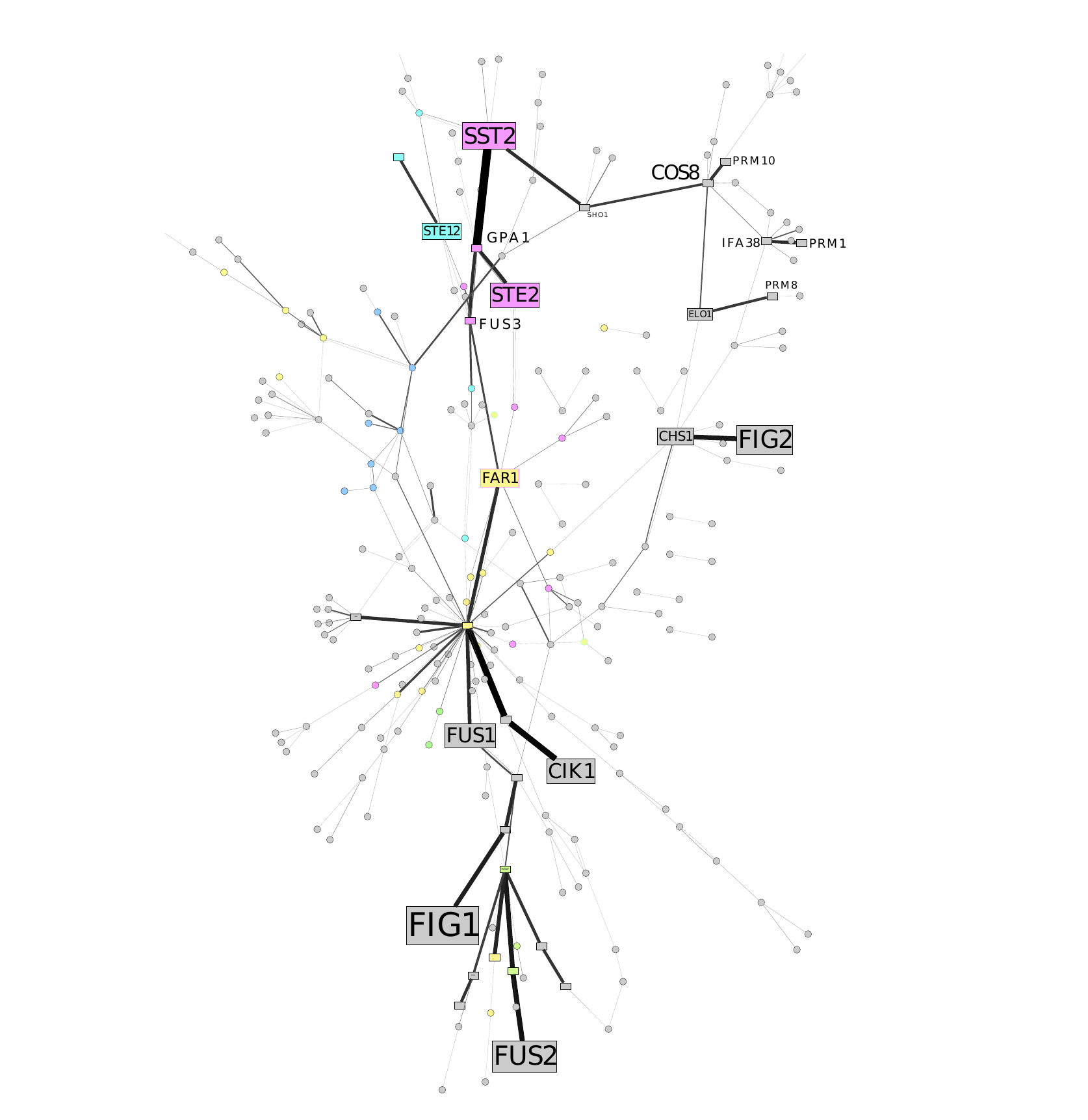}
\caption{This graph is a sub-network of the protein-protein interaction map,
obtained by including nodes that appear more than 30\% of the times on the 56
inferred Steiner trees for $\lambda=0.2$, with link intensity proportional to
the number of times the specific connection was found, and node size is
proportional to average prize. The layout was decided in order to minimize
crossings with the Graphviz suite. Afterward, colors were added denoting the
main GO annotation\label{network}: Actin (light green), Cell
Cycle (yellow), Chromatin structure (blue), Spindle Checkpoint (dark green),
Cell wall (cyan) and Pheromone sensing (magenta). The annotations were obtained
from the SGD project "Saccharomyces Genome Database"
\url{http://www.yeastgenome.org/}.  }
\end{figure}

The biological coherence of our results  was assessed by showing that the
average number of GO Slim annotations shared between neighbors in the inferred
trees is significantly superior to those of random trees ($t$-test,
$p<5.10^{-7}$, see SI).  
The results obtained from each one of the 56 expression patters have been
merged, leading to  a final network with links which are weighted by their
frequency of appearance. For each identified tree there is also a direction
associated to each link toward the chosen root (not displayed in Fig.
\ref{network} for clarity).

The network we found (see Fig~\ref{network}) contains known pathways, whose
completeness depends on $\lambda$. It shows signaling going from STE2, to the
cell cycle module (starting from CDC28), proteins involved in cell polarity and
cytoskeleton reorganization
(RVS161 and RVS167), and another branch containing essentially PRM proteins and
proteins involved in sphingolipid synthesis (ELO1) or cell wall (CHS1).  A
closer look at the structure of the inferred sub-network shows that it is
constituted of two types of proteins: 
proteins differentially expressed during pheromone response, already discovered
by transcriptomic studies, and protein bridging between sub-parts of the
network, while having a stable expression level in these conditions.
These proteins, that will be called herein {\it Steiner proteins}  in reference
to Steiner nodes in the MSTT, are not detected by classical transcriptomics, but
may have an informative or signaling role nonetheless: they allow signal
propagation between modules of the transduction network, and can be discovered
only through a combined analysis of the protein interaction and transcriptomics data.

One such protein we found is COS8, a gene of unknown function, which appear at
the head of the third main branch starting from STE2, linking between the
standard pheromone pathway and a group of proteins related to membrane
structure. To ensure that COS8 frequent occurrence is not due to a bias of our
optimization criterion to predilect proteins having high
connectivity\cite{White2008}, we assessed the statistical significance of the
Steiner proteins by bootstrap experiments with noisy costs and prizes.
Interestingly, the COS8 protein appears frequently enough to be considered as
significant and biologically relevant. Given the interest in uncovering a role
for a member of a large gene family in yeast (the COS family contains 11 highly
similar members), we therefore attempt to infer its biological role
using its neighborhood in the inferred trees.

COS8 is suggested to be the target of the STE2-GPA1-SST2-SHO1 cascade. As SHO1
is also a sensor involved in the filamentous growth and osmolarity pathway,
COS8 could be an actor of more than just the pheromone system, and could
provide cross-talks between these different pathways.  Out of 48 proteins
experimentally shown to interact with it, a small subset appear in most
simulations, composed of the membrane protein PRM10, the fatty acid elongases
ELO1, SUR4 and FEN1 -- involved in the first steps of sphingolipid
synthesis--, and other components of the secretion pathway and ceramide
synthesis such a AUR1, LAC1 and IFA38. Such homogeneity in the uncovered
interactions of COS8, relative to the diversity in the 48 interactants set,
allows us to putatively predict both localization and role of COS8:  it should
be an endoplasmic reticulum (ER) protein involved in sphingolipid synthesis. 
Concerning localization, these predictions precise the results of a previous
study using GFP fluorescence, that localized COS8 either in the ER or the
nucleus\cite{Spode2002}. 
Moreover, the single hint about COS8 function, uncovered in a large-scale
transcriptomic experiment, and not experimentally verified, is an undetermined
role in the unfolded protein response, a process occurring in the
ER\cite{Travers2000}, in agreement with our results.

As for COS8 role, sphingolipids are an essential components of the membrane,
being part of the lipid rafts microdomains \cite{Dickson2002}. To check the
relevance of COS8 for the sphingolipid synthesis, and more generally for
membrane structure, we analyzed with the same algorithm another large-scale
dataset \cite{Kuranda2006} testing for response to caffeine or rapamycin
stress, which are also known to inhibit TOR pathway \cite{Zheng1995} and
therefore to disrupt membrane structure. Consistently with the pheromone
results, COS8 was also detected as a Steiner protein, and was significantly
enriched in the resulting sub-graphs, with the same interacting partners. We
therefore decided to check experimentally our predictions about COS8
interactions and putative role.

\subsection{Experimental validation}
\begin{figure}
\includegraphics[width=1.0\columnwidth]{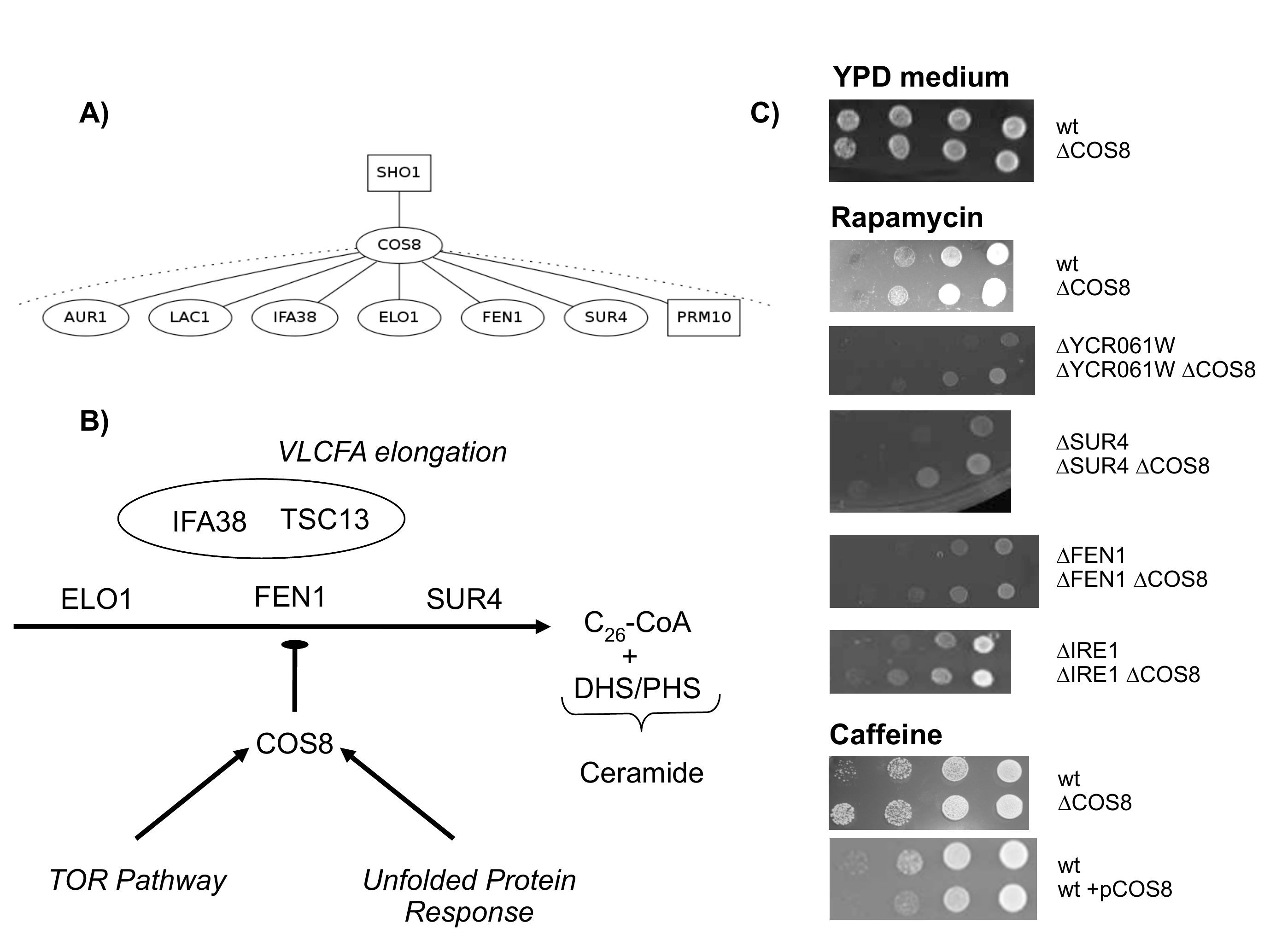}
\caption{A) The main proteins interacting with COS8 in our signaling tree.
Squares stand for membrane proteins. The dotted lines show that many protein
interactions are never found in our tree. B) A global scheme of the putative
negative regulatory role of COS8, at the interface between very long chain
fatty acid elongation (VLFCA), TOR signaling and the unfolded protein response
(See text). C) A subset of the genetic experiments showing i) the rescue of
$\Delta$YCR061W, $\Delta$SUR4, $\Delta$FEN1, and $\Delta$IRE1 by $\Delta$COS8
in rapamycin medium; ii) the effects of either the disruption or the
over-expression of COS8 in a caffein medium\label{experiments}}
\end{figure}

We proceeded to genetic experiments in strains with disrupted COS8 or
containing a plasmid over-expressing (Fig~\ref{experiments} ). We tested
interactions with the main components of sphingolipid synthesis, in various
conditions (full experimental details in SI text). 

We made the $\Delta$COS8 strain for this study, by replacing the chromosomal
copy of this gene by cos8::LEU2 deletion cassete. For the overexpression study,
we cloned COS8 gene in multicopy plasmid pRS426(with URA3 marker). To obtain
double mutants used in this study cos8::LEU2 deleted BY4741 strain was crossed
with a strain from YKO collection (ORF::KAN). Resulted diploid was dissected
and segregants having KAN and LEU2 markers were selected and used for further
experiments. Details concerning all the experiments are given in SI.

We found that deleting COS8 rescue completely the phenotype of $\Delta$FEN1 and
$\Delta$YCR061W strains in rapamycin medium -- as well as caffein --, and
partially the phenotype of $\Delta$SUR4 in the same conditions. Oppositely,  a
strain over-expressing COS8 is hypersensitive to caffein and presents the same
phenotype as FEN1 deletion. This shows a clear interaction between TOR
signaling, COS8 and long chain fatty acid elongation, and hints at a negative
regulation of very long chain fatty acid (VLCFA) elongation by COS8. 
As a control experiment, we checked for genetic interactions  with IRE1, the
master regulator of unfolded protein response, which is the annotated role of
COS8\cite{Travers2000}, and we also found this interaction (See
Fig.~\ref{experiments}). 
Finally, we tested for growth defect in $\Delta$COS8 strains  in media
containing myriocin. Myriocin is a very potent inhibitor of serine
palmitoyltransferase \cite{Miyake1995}, the first step in sphingosine
biosynthesis, and slow sphingolipid synthesis. We compared the sensitivity of
$\Delta$COS8 and wt strains to this antibiotic.  After 36 hours of incubation
in liquid YPD media with  0.1 mM  of myriocin we observed a transient but clear
positive effect of COS8 deletion on cell growth (See Table \ref{myriocin}). 

\begin{table}
\begin{center}
\begin{tabular}{ccc}
\hline
Strain & control & 0.1 mM myriocin \\
\hline
wt & 20 & 1.9 \\
$\Delta$COS8 & 20.8 & 6.7 \\
\hline
\end{tabular}
\caption{OD600 of the wt strain and the $\Delta$COS8 strain in myriocin and control medium.} \label{myriocin}
\end{center}
\end{table}

Weak but reproducible increase of myriocin resistance for $\Delta$COS8 could
also be seen on solid YPD media. This can be interpreted if COS8 indeed
regulates negatively the sphingolipid production, via VLCFA synthesis : in
these conditions the cell growth rate should become dependent on the efficiency
of VLCFA elongation, which is less restricted in $\Delta$COS8 strains,
therefore leading to smaller effects of myriocin.  We therefore conclude that
the function of COS8 is indeed related to sphingolipid metabolism, and probably
regulates negatively the VLCFA synthesis and therefore the sphingolipid
production.

\section{Discussion}

Algorithmic predictions and genetic experiments show interactions between
sphingolipid synthesis, particularly ceramide, pheromone response and TOR
signaling. Sphingolipids have recently been involved in the TOR-regulated
network\cite{Aronova2008, Mousley2008}: TOR is able to activate the
reaction of synthesis of ceramide from dihydrosphingosine. The molecular
mechanisms of this regulation are still unknown, but are coherent with our
experimental results about a negative role of COS8 in VLCFA elongation, as
$\Delta$COS8 strains are resistant to caffein and rapamycin -- two components
known to inhibit the TOR pathway. As sphingolipids are now discovered to be
both essential membrane components and signaling molecules, understanding
the regulation of their synthesis by various pathways, and the potential
cross-talk they could provide, is a crucial issue. Here, regulation of
sphingolipid synthesis by COS8 would provide the cell with the ability to
integrate signal from the pheromone pathway, the osmolarity pathway, and the
TOR pathway in order to modify its membrane structure. Finally, COS8 is a
member of a gene family of 11 highly similar members. Further investigations
would be needed to identify the functional role of other members of the family,
considering that both our predictions and experiments seem to indicate that
COS8 has a major effect, among all members of the COS family.

\section{Conclusions}
We have presented a new computational technique, inspired from statistical
physics, which  can efficiently extract new useful information about
interactions in signaling pathways (from gene expression and protein-protein
data) by solving an appropriately defined optimization problem on graphs. 
Our method not only provides candidate networks linking proteins of known
function; the method also suggests new roles for proteins of previously unknown
function. As a test case we specifically predict a functional role for the COS8
protein (a member of a large gene family with yet unknown functional role) both
in sphingolipid synthesis and in the TOR pathway in 
\emph{S. cerevisiae}. We have validated the prediction by providing
experimental evidence showing that COS8 is involved in a regulatory loop at the
level of ceramide synthesis.  Our new algorithmic technique has several
properties which should make it of significant value to optimization problems
in systems biology: efficiency (nearly linear time complexity), simplicity (it
is based on a fixed point equation), parallelizability, and the ability to
include other biological priors,
such as synthetic lethal interactions or phosphoproteomics data. 
Moreover, the technique outperforms the
best-known techniques in the field: we tested it on unsolved instances of the
best-known library (SteinLib), and it achieved better optima than were known
previously.  Finally, it is relatively easy to adapt the technique to a large
class of network reconstruction problems, including many which arise in systems
biology.

\begin{materials}
Full methods are in SI Text. They include: (1)  Algorithm design (the model,
derivation of the message-passing cavity equations, the max-sum limit,
computation of marginals, iterative dynamics and reinforcement, a note on
directness), (2) Numerical results on benchmark problems and direct  comparison
with LP-based techniques, (3) Data source and results (including data
concerning  GO annotation enrichment), (4) Experimental protocols (Strains,
media and culture conditions, Construction of multicopy plasmid with COS8
chromosomal allele, Construction of the COS8 deleted strain, Construction of
double mutants, Drug sensitivity assays), (5) Algorithm comparison with
previous data.

\end{materials}

\begin{acknowledgments}
This work has been supported by a Microsoft Research External Activities grant.
\end{acknowledgments}

\end{article}

\end{document}